\documentclass[aip,jcp,preprint]{revtex4-2}

\usepackage{amsmath}
\usepackage{booktabs}
\usepackage{graphicx}
\usepackage{listings}
\usepackage[version=4]{mhchem}
\usepackage[section]{placeins}
\usepackage{xcolor}

\newcommand*{\libpprpa}{\texttt{LibppRPA}}
\newcommand*{\pyscf}{\texttt{PySCF}}

\definecolor{codegreen}{rgb}{0,0.6,0}
\definecolor{codegray}{rgb}{0.5,0.5,0.5}
\definecolor{codepurple}{rgb}{0.58,0,0.82}
\definecolor{backcolour}{rgb}{0.95,0.95,0.92}

\lstdefinestyle{mystyle}{
    backgroundcolor=\color{backcolour},   
    commentstyle=\color{codegreen},
    keywordstyle=\color{magenta},
    numberstyle=\tiny\color{codegray},
    stringstyle=\color{codepurple},
    basicstyle=\ttfamily\footnotesize,
    breakatwhitespace=false,         
    breaklines=true,                 
    captionpos=b,                    
    keepspaces=true,                 
    numbers=left,                    
    numbersep=5pt,                  
    showspaces=false,                
    showstringspaces=false,
    showtabs=false,                  
    tabsize=1
}
\makeatother

\begin{document}

\title{\libpprpa{}: An Open-Source Library for Particle-Particle Random Phase Approximation}

\author{Jincheng Yu}
\thanks{These two authors contributed equally}
\affiliation{Department of Chemistry, Duke University, Durham, North Carolina 27708, United States}
\affiliation{Department of Chemistry and Biochemistry, University of Maryland, College Park, Maryland 20742, United States}
\author{Jiachen Li}
\thanks{These two authors contributed equally}
\affiliation{Department of Chemistry, Yale University, New Haven, Connecticut 06520, United States}
\affiliation{Department of Chemistry, Duke University, Durham, North Carolina 27708, United States}
\author{Chaoqun Zhang}
\affiliation{Department of Chemistry, Yale University, New Haven, Connecticut 06520, United States}
\author{Tianyu Zhu}
\affiliation{Department of Chemistry, Yale University, New Haven, Connecticut 06520, United States}
\author{Weitao Yang}
\email{weitao.yang@duke.edu}
\affiliation{Department of Chemistry, Duke University, Durham, North Carolina 27708, United States}


\begin{abstract}
The accurate description of electron correlation and excitation energies remains a fundamental challenge in quantum chemistry. The particle-particle random phase approximation (ppRPA) has emerged as a promising method for capturing a broad range of excited-state properties. 
However, the implementation of ppRPA has been largely limited to in-house software, restricting its accessibility and usability. In this work, we present \libpprpa{}, an open-source and lightweight Python library designed for efficient and flexible ppRPA calculations of (1) electronic excitation energy and its associated analytical gradients and (2) the ground state correlation energy, and its associated analytical gradients. \libpprpa{} enables seamless integration with existing quantum chemistry packages, such as \pyscf{}, by utilizing occupation numbers, molecular orbital coefficients, and three-center electron repulsion integrals. We implement both direct diagonalization and the iterative Davidson algorithm for solving the ppRPA equations, as well as active-space approximations, allowing users to balance accuracy and computational efficiency.
We demonstrate the performance of \libpprpa{} through benchmark calculations on singlet-triplet gaps, double excitations, charge-transfer excitations, and valence/Rydberg excitations, showcasing its reliability across diverse molecular systems. The library provides a robust platform for studying electronic excitations and offers new opportunities for future developments in electronic structure theory.
\end{abstract}

\maketitle

\section{Introduction}
The accurate description of electron correlation 
and electronic excitations is a central task 
in theoretical chemistry.
Over the past decades, 
significant efforts have been made to developing theoretical approaches 
that address these challenges,
leading to a variety of methods 
with distinct strengths and limitations.
Among the most widely used approaches 
are time-dependent density functional theory (TDDFT)\cite{
rungeDensityFunctionalTheoryTimeDependent1984,
casidaTimeDependentDensityFunctional1995,
ullrichTimeDependentDensityFunctionalTheory2011}, 
the Bethe-Salpeter equation (BSE) formalism\cite{
salpeterRelativisticEquationBoundState1951,
shamManyParticleDerivationEffectiveMass1966,
hankeManyParticleEffectsOptical1979}, 
and wave-function-based methods\cite{watsonBenchmarkingPerturbativeTripleExcitations2013,loosHowAccurateAre2021,verilQUESTDBDatabaseHighly2021,loosHowAccurateAre2021,j.bartlettPerspectiveCoupledclusterTheory2024}.

TDDFT has become one of the most popular computational approaches for describing molecular systems due to its balance of computational efficiency and accuracy.
The linear-response formulation of TDDFT has been widely implemented in modern quantum chemistry packages to calculate energies, structures, and other properties of excited states\cite{
jacqueminTDDFTPerformanceVisible2008,
casidaTimedependentDensityfunctionalTheory2009,
yuen-zhouTimeDependentDensityFunctional2010,
laurentTDDFTBenchmarksReview2013,
jinExcitedStateProperties2023,
knyshAssessingAccuracyTDDFT2024,
pohAlternantHydrocarbonDiradicals2024}.
With the iterative Davidson algorithm\cite{stratmannEfficientImplementationTimedependent1998},
the formal scaling of TDDFT is $\mathcal{O}(N^4)$,
where $N$ is the size of the system.
However,
its performance can suffer for charge-transfer (CT) states
and Rydberg states\cite{
tozerRelationshipLongrangeChargetransfer2003,
dreuwLongrangeChargetransferExcited2003}, 
partly due to the incorrect long-range behavior for describing the potential energy surface of 
TDDFT with conventional density functional approximations (DFAs).
Efforts to address these limitations have included 
the development of range-separated functionals\cite{
steinPredictionChargetransferExcitations2009,
refaely-abramsonFundamentalExcitationGaps2011}
and the tuning of 
the Hartree-Fock (HF) exchange fraction in DFAs\cite{
brucknerBenchmarkingSingletTriplet2017,
gongBenchmarkParameterTuning2020}. 
Furthermore, 
the performance of TDDFT is critically dependent on 
the selection of the exchange-correlation (XC) kernel\cite{
laurentTDDFTBenchmarksReview2013}, 
which plays a central role in determining its accuracy.

The BSE formalism\cite{
salpeterRelativisticEquationBoundState1951,
shamManyParticleDerivationEffectiveMass1966,
hankeManyParticleEffectsOptical1979},
derived from Green's function theory or many-body perturbation theory, 
has also gained growing attention for computing optical spectrum of molecules, interfaces and solids.
BSE is commonly performed on top of quasiparticle (QP) energies computed at the $GW$ level,
which is denoted as the BSE/$GW$ approach.
By accurately capturing the long-range behavior 
and using the dynamic screening interaction for non-local electron correlations in real systems, 
the BSE/$GW$ approach provides reliable predictions of 
excitation energies for a wide range of systems\cite{
blaseBetheSalpeterEquation2020,
moninoSpinConservedSpinFlipOptical2021,
jiangRealtimeGWBSEInvestigations2021,
knyshModelingExcitedState2022,
choSimplifiedGWBSE2022,
liCombiningRenormalizedSingles2022,
uddinAnisotropyOpticalProperties2022,
wuQuasiparticleExcitonicProperties2024,
bhattacharyaBSEGWPredictionCharge2024,
hillenbrandEnergyspecificBetheSalpeter2025,
zhouAllElectronBSEGWMethod2025,
liuManyBodyEffectsHeterogeneous2025}.
However,
there are several challenges for the BSE/$GW$ formalism.
First,
the accuracy of the BSE/$GW$ formalism heavily depends on 
the level of self-consistency in the $GW$ calculation\cite{
jacqueminAssessmentConvergencePartially2016,
hungExcitationSpectraAromatic2016,
liCombiningRenormalizedSingles2022,
forsterQuasiparticleSelfConsistentGWBetheSalpeter2022},
similar to the dependence on the DFA in TDDFT. 
Second, BSE/$GW$ suffers from the underscreening error, 
as a result of the missing vertex correction in the BSE kernel.
Third, although BSE has the same $\mathcal{O}(N^4)$ scaling as the TDDFT,
the proceeding $GW$ calculation can be computationally demanding.
Methods including obtaining QP energies from machine-learning frameworks\cite{venturellaMachineLearningManyBody2024,venturellaUnifiedDeepLearning2025} and combining BSE with generalized Kohn–Sham (KS) approaches, 
such as localized orbital scaling correction (LOSC)\cite{
liCombiningLocalizedOrbital2022}
and Koopmans-compliant functionals\cite{
elliottKoopmansMeetsBethe2019}, 
have been proposed to avoid the computational bottleneck. 
Additionally,
the analytic gradient of BSE/$GW$ excitation energies is not available,
where the relaxation of the excited-state structure can only be performed with approximations.

Alongside the computationally affordable linear-response formalisms, 
highly accurate wave function methods are commonly used to predict excited energies of molecular systems.
It has been shown that multireference methods such as complete active space second-order perturbation theory 
(CASPT2)\cite{anderssonSecondorderPerturbationTheory1992},
multireference configuration interaction (MRCI)\cite{siegbahnCompleteActiveSpace1981}
can predict different excited states on the equal footing.
However,
the selection of the active space in multireference methods can be ambiguous.
Another path is using single-reference wave function methods such as 
algebraic diagrammatic construction\cite{vonniessenComputationalMethodsOneparticle1984}
and coupled cluster techniques\cite{
kochExcitationEnergiesCoupled1990}. 
These methods offer a systematically improvable path to increase the accuracy by including high-order expansions.
However, 
due to their high computational cost, 
wave function methods are typically reserved for 
benchmark studies\cite{loosMountaineeringStrategyExcited2018,
loosMountaineeringStrategyExcited2020,
verilQUESTDBDatabaseHighly2021}.

Recently the particle-particle random phase approximation (ppRPA) has gained increasing attentions
for ground-state and excited-state properties of molecules and solid-state materials.
ppRPA was originally developed to calculate
the nuclear many-body correlation energy\cite{
ripkaQuantumTheoryFinite1986,
ringNuclearManyBodyProblem2004},
and has been extended to describe correlation in electronic systems by the Yang group \cite{vanaggelenExchangecorrelationEnergyPairing2013,vanaggelenExchangecorrelationEnergyPairing2014}. 
It can be derived from different approaches,
including the adiabatic connection
using the pairing matrix fluctuations\cite{
vanaggelenExchangecorrelationEnergyPairing2013,
vanaggelenExchangecorrelationEnergyPairing2014}
TDDFT with the pairing field\cite{
pengLinearresponseTimedependentDensityfunctional2014},
and the equation of motion (EOM)\cite{
roweEquationsofMotionMethodExtended1968,ringNuclearManyBodyProblem2004},
and particle-particle channel in BSE\cite{marieAnomalousPropagatorsParticleparticle2024}.
Compared with 
the particle-hole random phase approximation 
(phRPA)\cite{bohmCollectiveDescriptionElectron1951,
renRandomphaseApproximationIts2012},
ppRPA contains information in 
the particle-particle 
and the hole-hole channels,
which associates with the fluctuation of the pairing matrix\cite{vanaggelenExchangecorrelationEnergyPairing2013}.
For the ground-state energy,
ppRPA is equivalent to ladder couple cluster doubles when the Hartree-Fock approximation is used for the nonintereting reference system (CCD)\cite{pengEquivalenceParticleparticleRandom2013,scuseriaParticleparticleQuasiparticleRandom2013},
which is exact up to the second order with the ladder partial summation. 
Recently,
the extension of ppRPA based on a multireference character has been developed to describe the strong correlation in molecular systems\cite{wangUnifiedDiagrammaticFormulation2025}.
ppRPA has also been extended to obtain charge-neutral excitation energies by the Yang group \cite{yangDoubleRydbergCharge2013,berkelbachCommunicationRandomphaseApproximation2018}:
it calculates the two-electron addition and 
the two-electron removal energies,
which can be considered as an approximation to double-electron-affinity or double-ionization-potential EOM-CCD.
By taking the differences between 
the two-electron addition energies of 
the ($N-2$)-electron system\cite{yangDoubleRydbergCharge2013}
or the two-electron removal energies of 
the ($N+2$)-electron system\cite{yangDoubleRydbergCharge2013, bannwarthHoleHoleTamm2020,
yuInitioNonadiabaticMolecular2020},
excitation energies of the $N$-electron system can be obtained.

ppRPA has gained great success in describing 
various ground-state and excited-state properties.
For example,
ppRPA can accurately predict a variety types of excitation energies including 
valence excitations\cite{yangExcitationEnergiesParticleparticle2014}, 
singlet-triplet (S-T) energy gaps of diradicals\cite{yangSingletTripletEnergy2015,yangNatureGroundElectronic2016},
charge-transfer (CT) excitations\cite{
yangChargeTransferExcitations2017}, 
double excitations\cite{yangDoubleRydbergCharge2013,
yuAccurateEfficientPrediction2025},
and Rydberg excitations\cite{yangDoubleRydbergCharge2013}.
Conical intersections\cite{yangConicalIntersectionsParticle2016}, oscillator strengths\cite{yangDoubleRydbergCharge2013}, 
and excited-state geometries\cite{zhangAnalyticGradientsGeometry2014} can also be well described by ppRPA.
More recently, 
with the development of the active-space formalism\cite{
liLinearScalingCalculations2023,zhangAccurateEfficientCalculation2016},
the computational cost of ppRPA can be significantly reduced without loss of accuracy.
With this development, 
ppRPA has also been applied to predict accurate excitation energies
of point defects\cite{
liAccurateExcitationEnergies2024,
liParticleParticleRandom2024}.
In addition,
ppRPA with the Tamm-Dancoff approximation (TDA) has been applied
in the multireference DFT to predict 
dissociation energies and excitation energies\cite{
chenMultireferenceDensityFunctional2017,
liMultireferenceDensityFunctional2022}.
In the context of the Green's function theory,
the ppRPA eigenstates are also utilized in the T-matrix self-energy 
to calculate QP energies\cite{zhangAccurateQuasiparticleSpectra2017,liRenormalizedSinglesGreens2021,loosStaticDynamicBethe2022},
as a counterpart of phRPA used to construct the $GW$ self-energy,

The success of ppRPA can be attributed to several reasons.
First,
ppRPA can be viewed as an embedding approach in the Fock space.
Two frontier electrons are treated in an exact way 
using subspace CI with a seamless integration 
of DFT for the remaining ($N-2$) electrons\cite{
yangNatureGroundElectronic2016,
zhangAccurateEfficientCalculation2016}.
Thus, for diradicals and a large number point defects,
ppRPA describes the strong correlation of the two added electrons, 
free from the static correlation error and the spin contamination.
Second,
ppRPA contains the two-particle information by its construction.
Therefore, 
ppRPA is naturally capable of describing double excitations\cite{
yangDoubleRydbergCharge2013,
yuAccurateEfficientPrediction2025},
which cannot be captured by linear-response particle-hole formalisms such as TDDFT and BSE with the adiabatic approximation.
Third,
ppRPA kernel presents the correct long-range asymptotic behavior,
so it can accurately predict CT and Rydberg excitations\cite{
yangDoubleRydbergCharge2013,
yangChargeTransferExcitations2017}.
Fourth,
because the ppRPA kernel is independent of the parent DFA and excitation energies in ppRPA are obtained as the difference two-electron addition/removal energies,
the starting DFA dependence of ppRPA is small compared with 
TDDFT\cite{
yangExcitationEnergiesParticleparticle2014,
yangChargeTransferExcitations2017},
whose accuracy largely depends on the fundamental gap at the DFT level and the form of the XC kernel.

In this work,
we develop a reliable and stable implementation for ppRPA
in \libpprpa{},
which will serve as an open-source and lightweight library
conducting ppRPA calculations for electronic and magnetic properties of molecular and periodic systems ($\Gamma$-point supercell approach).
With the implementation completely in Python,
we aim to provide \libpprpa{} users with great flexibility 
to perform ppRPA calculations based on data 
from any quantum chemistry packages of their choice.
To demonstrate this
we integrate \libpprpa{} with \pyscf{}\cite{sunPySCFPythonbasedSimulations2018,sunRecentDevelopmentsPySCF2020}.
In principle,
\libpprpa{} can take input generated from any Gaussian type orbital (GTO) based software,
which provides immediate access for researchers
to perform calculations with ppRPA.

\section{Theory}

\subsection{Pairing Matrix}
The ppRPA formalism can be derived from different approaches,
including the equation of motion\cite{ringNuclearManyBodyProblem2004,
roweEquationsofMotionMethodExtended1968}, 
the adiabatic connection\cite{vanaggelenExchangecorrelationEnergyPairing2013,
vanaggelenExchangecorrelationEnergyPairing2014}, 
TDDFT with the pairing field\cite{pengLinearresponseTimedependentDensityfunctional2014},
and the BSE in the Green's function theory\cite{marieAnomalousPropagatorsParticleparticle2024}.
As a counterpart to phRPA formulated with the fluctuation of the density-density response function,
the ppRPA formalism is formulated with the fluctuation of the pairing matrix\cite{vanaggelenExchangecorrelationEnergyPairing2013,vanaggelenExchangecorrelationEnergyPairing2014,ringNuclearManyBodyProblem2004,ripkaQuantumTheoryFinite1986}
\begin{equation}
    \kappa (x_1, x_2, t) = \left \langle 
    \Psi^N_0 \left | \hat{\psi} (x_2, t) \hat{\psi} (x_1, t) \right | \Psi^N_0 
    \right \rangle
\end{equation}
where $x=(r,\sigma)$ is the combined space-spin variable, 
$\Psi^N_0$ is the $N$-electron ground state, 
$\hat{\psi}^{\dagger}$ and $\hat{\psi}$ are the second quantization creation and annihilation operator.
In the absence of an external pairing field,
the pairing matrix $\kappa (x_1, x_2, t)$ is zero, 
but its linear response to an external pairing field $\delta \kappa (x_1, x_2, t)$ is not zero and is described with the retarded dynamic pairing matrix fluctuation $\bar{K}(t-t')$ is\cite{vanaggelenExchangecorrelationEnergyPairing2013}
\begin{equation}
    \bar{K}_{pqrs}(t-t') = -i \theta(t-t')
    \left \langle
    \Psi^N_0 | 
    [ \hat{a}_p (t) \hat{a}_q (t) , 
    \hat{a}^{\dagger}_s (t') \hat{a}^{\dagger}_r (t')] 
    | \Psi^N_0
    \right \rangle
\end{equation}
which has poles at two-electron addition and removal energies that characterize double electron addition and ionization processes in the frequency space
\begin{equation}\label{eq:retarded_fluctuation_frequency}
    \bar{K}(\omega)_{pqrs} = 
    \sum_m \frac{\left \langle \Psi^N_0 | \hat{a}_p \hat{a}_q | \Psi^{N+2}_0 \right \rangle 
    \left \langle \Psi^{N+2}_0 | \hat{a}_s^{\dagger} \hat{a}_r^{\dagger} | \Psi^N_0 \right \rangle }
    {\omega - \Omega^{N+2}_m + i\eta }
    - \sum_m \frac{\left \langle \Psi^N_0 | \hat{a}_s^{\dagger} \hat{a}_r^{\dagger} | \Psi^{N-2}_0 \right \rangle \left \langle \Psi^{N-2}_0 | \hat{a}_p \hat{a}_q | \Psi^N_0 \right \rangle }
    {\omega - \Omega^{N-2}_m + i\eta }
\end{equation}
Here  $\hat{a}_p^{\dagger}$ and $\hat{a}_p$ are the second quantization creation and annihilation operator for the orbital $p$, 
$\Omega^{N\pm2}$ is the two-electron addition/removal energy, 
$\eta$ is a positive infinitesimal number.
We use $i$, $j$, $k$, $l$ for occupied orbitals, 
$a$, $b$, $c$, $d$ for virtual orbitals, 
$p$, $q$, $r$, $s$ for general orbitals, 
and $m$ for the index of the two-electron addition/removal energy.  

In many-body perturbation theory, 
the time-ordered form is normally used for theoretical derivation. 
The time-ordered dynamic pairing matrix fluctuation is\cite{vanaggelenExchangecorrelationEnergyPairing2013}
\begin{equation}
    K_{pqrs}(t-t') =
    -i \left \langle
    \Psi^N_0 | 
    T [ \hat{a}_p (t) \hat{a}_q (t) , 
    \hat{a}^{\dagger}_s (t') \hat{a}^{\dagger}_r] (t')
    | \Psi^N_0
    \right \rangle 
\end{equation}
and in the frequency space is
\begin{equation}\label{eq:time_ordered_fluctuation_frequency}
    K(\omega)_{pqrs} = 
    \sum_m \frac{\left \langle \Psi^N_0 | \hat{a}_p \hat{a}_q | \Psi^{N+2}_0 \right \rangle 
    \left \langle \Psi^{N+2}_0 | \hat{a}_s^{\dagger} \hat{a}_r^{\dagger} | \Psi^N_0 \right \rangle }
    {\omega - \Omega^{N+2}_m + i\eta }
    - \sum_m \frac{\left \langle \Psi^N_0 | \hat{a}_s^{\dagger} \hat{a}_r^{\dagger} | \Psi^{N-2}_0 \right \rangle \left \langle \Psi^{N-2}_0 | \hat{a}_p \hat{a}_q | \Psi^N_0 \right \rangle }
    {\omega - \Omega^{N-2}_m - i\eta }
\end{equation}

The dynamic paring matrix fluctuation, as shown in Eq.\ref{eq:time_ordered_fluctuation_frequency} clearly contains the double electron addition and removal excitation energies. 
It has also been shown that the dynamic paring matrix fluctuation provides a rigorous formulation for electron correlation in the particle-particle channel \cite{vanaggelenExchangecorrelationEnergyPairing2013,
vanaggelenExchangecorrelationEnergyPairing2014}, 
parallel to the electron correlation formulation in the particle-hole channel through the dynamic density density fluctuation, 
or the dynamic density response function\cite{langrethExchangecorrelationEnergyMetallic1977}.

To obtain the pairing matrix fluctuation $K$ of the interacting system,
the ppRPA approximates $K$ in terms of the non-interacting $K_0$ by the Dyson equation\cite{vanaggelenExchangecorrelationEnergyPairing2013,vanaggelenExchangecorrelationEnergyPairing2014}
\begin{equation}\label{eq:dyson}
    K = K^0 + K^0 V K
\end{equation}
where the antisymmetrized interaction $V_{pqrs}=\langle pq || rs \rangle = \langle pq | rs \rangle - \langle pq | sr \rangle$ with $\langle pq | rs \rangle = \int d\mathbf{x} d\mathbf{x'} \frac{\psi^*_p(\mathbf{x}) \psi_r(\mathbf{x}) \psi^*_q(\mathbf{x'}) \psi_s(\mathbf{x'})}{|\mathbf{r}-\mathbf{r'}|}$.
In Ref.\citenum{marieAnomalousPropagatorsParticleparticle2024},
the screened interaction is used in the Dyson equation,
which leads to the BSE in the particle-particle channel.

\subsection{ppRPA Equation}
The Dyson equation in Eq.\ref{eq:dyson} can be cast into a generalized eigenvalue problem that gives two-electron addition and removal energies\cite{vanaggelenExchangecorrelationEnergyPairing2013,yangExcitationEnergiesParticleparticle2014}
\begin{equation}\label{eq:eigen_equation}
\begin{bmatrix}\mathbf{A} & \mathbf{B} \\
\mathbf{B}^{\dagger}      & \mathbf{C}
\end{bmatrix}
\begin{bmatrix}
\mathbf{X} \\
\mathbf{Y}
\end{bmatrix}
=\Omega
\begin{bmatrix}
\mathbf{I} & \mathbf{0}  \\
\mathbf{0} & \mathbf{-I}
\end{bmatrix}
\begin{bmatrix}
\mathbf{X} \\
\mathbf{Y}
\end{bmatrix}
\end{equation}
with
\begin{align}
A_{ab,cd} & =\delta_{ac}\delta_{bd}(\epsilon_{a}+\epsilon_{b})+\langle ab||cd\rangle \\
B_{ab,kl} & =\langle ab||kl\rangle \\
C_{ij,kl} & =-\delta_{ik}\delta_{jl}(\epsilon_{i}+\epsilon_{j})+\langle ij||kl\rangle 
\end{align}
where $a<b$, $c<d$, $i<j$, $k<l$, 
the eigenvalue $\Omega$ is the two-electron addition/removal energy,
$X$ and $Y$ are the two-electron addition and removal eigenvectors.

For closed-shell systems,
the ppRPA equation in Eq.~\ref{eq:eigen_equation} can be cast into the spin-adapted form\cite{yangBenchmarkTestsSpin2013}.
For singlet excitations the ppRPA matrix elements are
\begin{align}
A^{\text{s}}_{ab,cd} & =\delta_{ac}\delta_{bd}(\epsilon_{a}+\epsilon_{b}) 
+ \frac{1}{\sqrt{(1+\delta_{ab})(1+\delta_{cd})}} (\langle ab|cd\rangle + \langle ab|dc\rangle) \\
B^{\text{s}}_{ab,kl} & = \frac{1}{\sqrt{(1+\delta_{ab})(1+\delta_{kl})}} 
(\langle ab|kl\rangle + \langle ab|lk\rangle) \\
C^{\text{s}}_{ij,kl} & =-\delta_{ik}\delta_{jl}(\epsilon_{i}+\epsilon_{j})
+ \frac{1}{\sqrt{(1+\delta_{ij})(1+\delta_{kl})}} (\langle ij|kl\rangle + \langle ij|lk\rangle)
\end{align}
with $a \leq b$, $c \leq d$, $i \leq j$ and $k \leq l$.
And for triplet excitations the ppRPA matrix elements are
\begin{align}
A^{\text{t}}_{ab,cd} & =\delta_{ac}\delta_{bd}(\epsilon_{a}+\epsilon_{b})+\langle ab||cd\rangle \\
B^{\text{t}}_{ab,kl} & =\langle ab||kl\rangle \\
C^{\text{t}}_{ij,kl} & =-\delta_{ik}\delta_{jl}(\epsilon_{i}+\epsilon_{j})+\langle ij||kl\rangle
\end{align}
with  $a<b$, $c<d$, $i<j$ and $k<l$.

ppRPA eigenvalues obtained from Eq.\ref{eq:eigen_equation} can be considered as an approximation to double-electron-affinity or
double-ionization-potential equation-of-motion coupled-cluster doubles\cite{yangDoubleRydbergCharge2013,berkelbachCommunicationRandomphaseApproximation2018}.
The ppRPA eigenvector is normalized as\cite{vanaggelenExchangecorrelationEnergyPairing2013,vanaggelenExchangecorrelationEnergyPairing2014}
\begin{equation} \label{eq:xy_norm}
    X^{m,\dagger} X^m - Y^{m,\dagger} Y^m = \pm 1
\end{equation}
where the upper sign is for two-electron addition excitations and the lower sign is for two-electron removal excitations.

The neutral excitation energy of an $N$-electron system can be obtained from the energy difference between two-electron addition energies of the corresponding ($N-2$)-electron system,
or the energy difference between two-electron removal energies of the corresponding ($N+2$)-electron system.
Similar to TDDFT and BSE,
the scaling of solving the ppRPA equation in Eq.\ref{eq:eigen_equation} is $\mathcal{O}(N^4)$ with the Davidson algorithm\cite{yangExcitationEnergiesParticleparticle2014}.
In the recently developed active-space formalism,
ppRPA excitation energies are shown to converge rapidly with canonical active-space orbitals for both molecular and periodic systems\cite{liLinearScalingCalculations2023,liAccurateExcitationEnergies2024,liParticleParticleRandom2024},
which significantly lowers the computational cost.

\subsection{ppRPA for Fractional Charge Systems}
With the extension of the Green's function theory to fractional charge systems established in Ref.\citenum{yangExtensionManybodyTheory2013},
the ppRPA equation in Eq.\ref{eq:eigen_equation} can be applied for fractional charge system with matrix elements defined as
\begin{align}
A_{ab,cd} & =\delta_{ac}\delta_{bd}(\epsilon_{a}+\epsilon_{b}) + 
\sqrt{(1-n_a)(1-n_b)(1-n_c)(1-n_d)} \langle ab||cd\rangle \\
B_{ab,kl} & = \sqrt{(1-n_a)(1-n_b)n_k n_k} \langle ab||kl\rangle \\
C_{ij,kl} & =-\delta_{ik}\delta_{jl}(\epsilon_{i}+\epsilon_{j}) + 
\sqrt{n_i n_j n_k n_k} \langle ij||kl\rangle 
\end{align}
where $n$ is the occupation number.
The total energy from Hartree-Fock energy with ppRPA correlation energy shows a linear energy behavior for systems with fractional charges and meets the flat-plane condition\cite{vanaggelenExchangecorrelationEnergyPairing2013}.
The fractional formulation of ppRPA also allows one to obtain the ppRPA chemical potential with the finite difference approach,
which significantly outperforms phRPA for predicting ionization potentials and electron affinity\cite{vanaggelenExchangecorrelationEnergyPairing2013}.

\subsection{ppRPA for Relativistic Hamiltonian}
Li and co-workers have extended particle-particle TDA method to treat relativistic two-component Hamiltonian \cite{williams-youngRelativisticTwoComponentParticle2016}. 
ppRPA for relativistic Hamiltonian is also implemented in \libpprpa{}, compatible with relativistic generalized mean-field reference using spin orbital basis functions or j-adapted spinors.
Various exact two-component theory (X2C) \cite{dyallInterfacingRelativisticNonrelativistic1997,kutzelniggQuasirelativisticTheoryEquivalent2005,iliasInfiniteorderTwocomponentRelativistic2007,liuExactTwocomponentHamiltonians2009} Hamiltonians are supported, such as the X2C theory in its one-electron variant (X2C-1e) and the X2C-1e with atomic mean-field two-electron interaction (the X2CAMF scheme) \cite{hessMeanfieldSpinorbitMethod1996,liuAtomicMeanfieldSpinorbit2018,zhangAtomicMeanfieldApproach2022}.
Two-component ppRPA calculations can be run seamlessly using the interface to \pyscf{}.
We note that the relativistic Hamiltonian is treated variationally, i.e., added at the mean-field level. 
With the no-pair approximation, the ppRPA working equations remain the same as those for the non-relativistic calculations except for the lack of spin symmetry.

\subsection{Total Energy}
The ppRPA correlation energy can be obtained in terms of the solution to generalized eigenvalue problem in Eq.\ref{eq:eigen_equation} as\cite{vanaggelenExchangecorrelationEnergyPairing2013,vanaggelenExchangecorrelationEnergyPairing2014}
\begin{equation}\label{eq:correlation_energy}
    E^{\text{pp,c}} = \sum \Omega^{\text{+2e}} - \textbf{Tr} \mathbf{A} 
    = - \sum \Omega^{\text{-2e}} - \textbf{Tr} \mathbf{C}
\end{equation}
where $\textbf{Tr}$ means trace,
$\Omega^{\pm\text{2e}}$ is two-electron addtion/removal energy.
The expression in Eq.\ref{eq:correlation_energy} is the same for fractional charge systems.
For ground states,
the ppRPA correlation energy can alternatively be interpreted as
the sum of all ladder diagrams in the wavefunction theory,
which is equivalent to the ladder-coupled-cluster doubles\cite{pengEquivalenceParticleparticleRandom2013,scuseriaParticleparticleQuasiparticleRandom2013}.
Then the ppRPA total energy is expressed as
\begin{equation}
    E^{\text{pp}} = E^{\text{HF}} + E^{\text{pp,c}}
\end{equation}
where $E^{\text{HF}}$ is the Hartree-Fock energy evaluated with the input orbitals.

Alternatively,
the ppRPA total energy can be obtained from the multireference DFT method\cite{chenMultireferenceDensityFunctional2017,liMultireferenceDensityFunctional2022},
which means
\begin{equation}\label{eq:mr}
    E^{\text{pp,MR}}_m = E^{\text{DFT,}N\pm2} + \Omega^{\mp2e}_m
\end{equation}
where $E^{\text{DFT,}N\pm2}$ is the DFT energy of the ($N\pm2$)-electron system and $\Omega^{\pm2e}$ is the two-electron addition/removal energy from ppRPA.
Eq.\ref{eq:mr} describes both ground and excited states on equal footing.
As shown in Refs.\citenum{chenMultireferenceDensityFunctional2017,liMultireferenceDensityFunctional2022},
the SCF solution of the multireference ppRPA energy can be achieved with the (generalized) optimized effective potential method\cite{yangDirectMethodOptimized2002,jinGeneralizedOptimizedEffective2017}.

\subsection{Natural Transition Orbital}
The natural transition orbitals (NTOs) of ppRPA is developed to provide qualitative descriptions of electronic transitions\cite{liParticleParticleRandom2024}.
For the NTOs of particle-hole formalisms,
the dominant particle-hole pairs of an excited state are obtained from the SVD of the corresponding transition density matrix\cite{martinNaturalTransitionOrbitals2003}.
Similarly, 
NTOs in ppRPA convey information about particle-particle pairs and hole-hole pairs.

For the $m$-th state,
the two-electron addition eigenvector $X^m$ can be viewed as a triangular matrix of dimension $N_{\text{vir}}\times N_{\text{vir}}$ and the two-electron removal eigenvector $Y^m$ can be viewed as a triangular matrix with of dimension $N_{\text{occ}}\times N_{\text{occ}}$.
Therefore,
the coefficients that transform molecular orbitals to natural transition orbitals of two particles (holes) can be obtained with the SVD of $X^m$ ($Y^m$)
\begin{align}
    X^m = & C^{\text{p1}, m} \sqrt{\lambda^{\text{p}, m}} C^{\text{p2}, m\dagger} \label{eq:nto_pp} \\
    Y^m = & C^{\text{h1}, m} \sqrt{\lambda^{\text{h}, m}} C^{\text{h2}, m\dagger} \label{eq:nto_hh}
\end{align}
In Eq.~\ref{eq:nto_pp},
the NTO coefficient matrices $C^{\text{p1}}$ and $C^{\text{p2}}$ of dimension $N_{\text{vir}}\times N_{\text{vir}}$ are associated with particle-particle pairs for adding two electrons,
which are weighted with diagonal elements of matrix $\lambda^{\text{p}}$.
Similarly,
in Eq.~\ref{eq:nto_hh},
the NTO coefficient matrices $C^{\text{h1}}$ and $C^{\text{h2}}$ of dimension $N_{\text{occ}}\times N_{\text{occ}}$ are associated with hole-hole pairs for removing the two electrons,
which are weighted with diagonal elements of matrix $\lambda^{\text{h}}$.
As a consequence of the normalization in Eq.~\ref{eq:xy_norm},
the NTO weights satisfy the following relation
\begin{equation}
    \sum_a^{\text{vir}} \lambda^{\text{p}}_a - \sum_i^{\text{occ}} \lambda^{\text{h}}_i = \pm 1
\end{equation}
where the upper sign is for two-electron addition excitations and the lower sign is for two-electron removal excitations. 
The resulting NTO weights ($\lambda_a^{\text{p}}$ and $\lambda_i^{\text{h}}$) can thus be employed to qualitatively analyze the components and multireference character of the associated ground and excited states.

\subsection{Density Matrix}
As shown in Ref.\citenum{jinIntroductoryLectureWhen2020},
the density of a DFA can be obtained as the functional derivative of the energy with respect to the external potential
\begin{equation}
    \rho (r) = \frac{\delta E}{\delta v_{\text{ext}} (r)}
\end{equation}
which is used for deriving phRPA and ppRPA density\cite{jiangRealtimeGWBSEInvestigations2021}.
Here we generalize the expression from the density to the one-particle density matrix,
whose occupied-occupied and the virtual-virtual blocks in the orbital space are defined as
\begin{align}
    D_{ij} & = \langle \Psi | \hat{a}_i \hat{a}_j^{\dagger} | \Psi \rangle \\
    D_{ab} & = \langle \Psi | \hat{a}_a^{\dagger} \hat{a}_b | \Psi \rangle
\end{align}

In ppRPA,
the energy is obtained as the summation of the energy of the ($N\pm2$)-electron reference system calculated by DFT and the two-electron addition/removal energy calculated by ppRPA.
Therefore, 
the total density matrix of the $m$-th state of the $N$-electron system is divided into two parts:
the non-interacting part of the ($N\pm2$)-electron reference system,
and the two-electron addition/removal part
\begin{equation}\label{eq:total_dm}
    D^{N,m} = D^{N\pm 2,\text{KS}} + D^{\text{2e}, m}
\end{equation}
which has a zero occupied-virtual block and agrees with the ppRPA density derived in Ref.\citenum{jinIntroductoryLectureWhen2020} in the real-space diagonal limit.

In Eq.\ref{eq:total_dm},
the two-electron addition/removal part is constructed with the ppRPA eigenvectors solved from Eq.\ref{eq:eigen_equation},
which contains information of adding and removing two electrons.
For the $m$-th excitation,
the virtual-virtual block for adding two electrons is
\begin{equation}
    D^{\text{2e}, m}_{ab} = [X^m X^{m,\dagger}]_{ab} + [X^{m,\dagger} X^m]_{ab} 
\end{equation}
and the occupied-occupied block for removing two electrons is
\begin{equation}
    D^{\text{2e}, m}_{ij} = - [Y^m Y^{m,\dagger}]_{ij} - [Y^{m,\dagger} Y^m]_{ij}
\end{equation}
Then the two-electron addition/removal density matrix stratifies the following relation
\begin{equation}\label{eq:dm_constrain}
    \textbf{Tr} [ D^{\text{2e}, m} ] = \pm 2
\end{equation}
where the upper sign is for two-electron addition excitations and the lower sign is for two-electron removal excitations in Eq.~\ref{eq:dm_constrain}.

\subsection{Analytic Gradient}
To obtain the energy gradient in ppRPA,
the total energy is expressed in the multireference DFT manner by Eq.\ref{eq:mr},
then the gradient can be calculated as\cite{zhangAnalyticGradientsGeometry2014}
\begin{equation}
    \frac{\partial E^{\text{pp,MR}}_m}{\partial \lambda} 
    = \frac{\partial E^{\text{DFT,}N\pm2}}{\partial \lambda} 
    + \frac{\partial \Omega^{\mp2e}_m}{\partial \lambda} 
\end{equation}
where $\lambda$ is the nuclear coordinate.
By virtue of the Hellmann–Feynman theorem,
the gradient of the two-electron addition/removal energy with respect to the nuclear coordinate is\cite{zhangAnalyticGradientsGeometry2014}
\begin{equation}
    \frac{\partial \Omega^{\mp2e}_m}{\partial \lambda} =
    \begin{bmatrix}
        X_m^\dagger \ Y_m^\dagger
    \end{bmatrix}
    \left (
    \frac{\partial}{\partial \lambda}
    \begin{bmatrix}
        \mathbf{A} & \mathbf{B} \\
        \mathbf{B}^{\dagger}      & \mathbf{C}
    \end{bmatrix}
    \right )
    \begin{bmatrix}
        X_m \\
        Y_m
    \end{bmatrix}
\end{equation}
where more details can be found in Ref.\citenum{zhangAnalyticGradientsGeometry2014}.

\section{Implementation Details}
\begin{figure}
\includegraphics[width=0.6\textwidth]{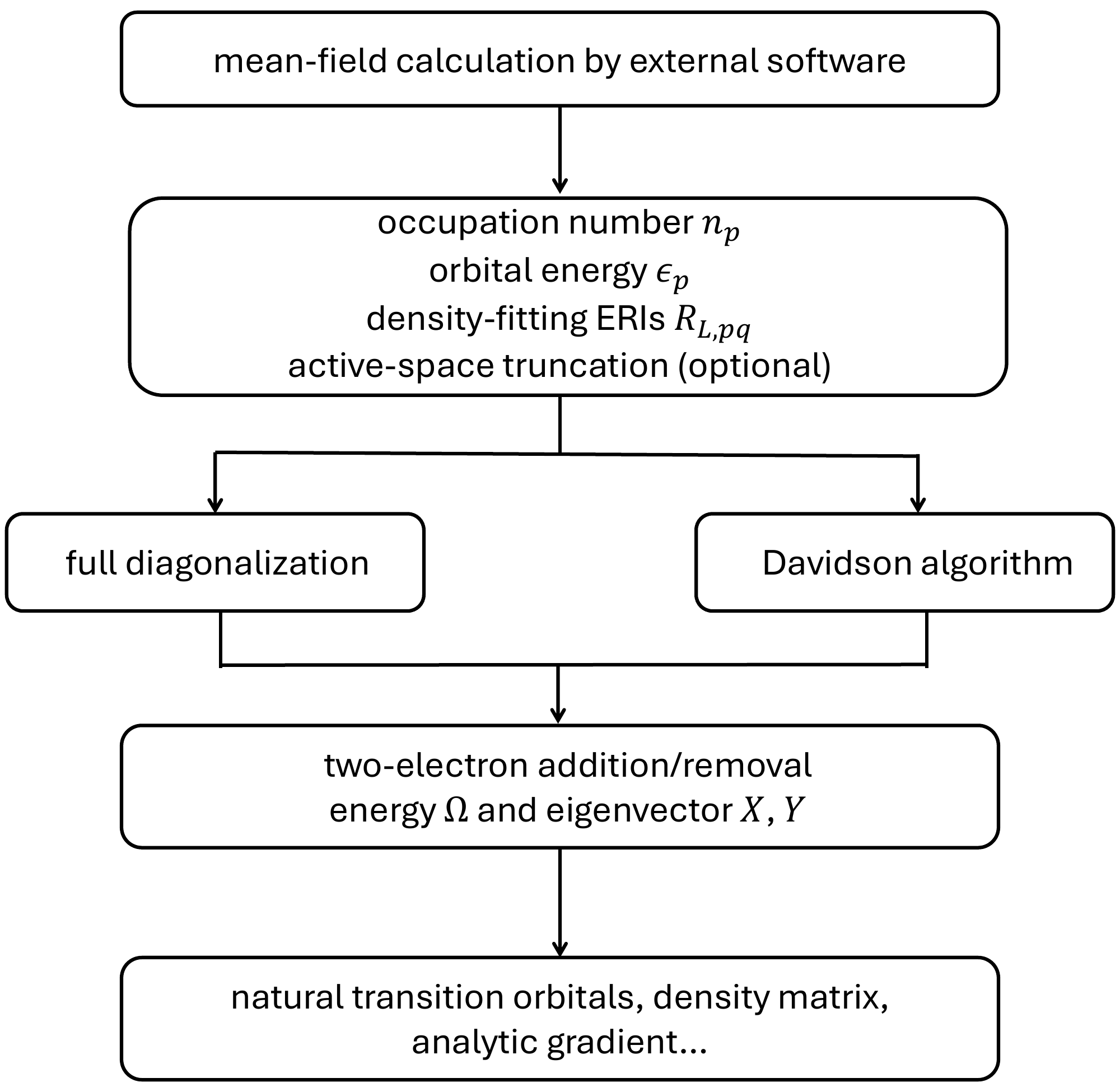}
\caption{The architecture of \libpprpa{}.}
\label{fig:workflow}
\end{figure}

\FloatBarrier
The \libpprpa{} library is an open-source and pure-Python package for performing ppRPA calculations within the Gaussian type orbital (GTO) framework.
The workflow is illustrated in Fig.\ref{fig:workflow}.
\libpprpa{} takes three main input variables:
a) occupation numbers, 
b) molecular orbital (MO) energies
and c)the three-center Gaussian density-fitting integrals in the MO space,
which can be from either a molecular or periodic (supercell with $\Gamma$-point sampling) calculation.
The input variables are passed to \libpprpa{} in the numpy.ndarray format, 
which can be generated on the fly or read via HDF5 binary data file from popular quantum chemistry software and in-house programs.
\libpprpa{} also offers a convenient interface that directly initializes the ppRPA calculation from a mean-field calculation by \pyscf{}\cite{sunPySCFPythonbasedSimulations2018,sunRecentDevelopmentsPySCF2020}.

After loading the input variables,
\libpprpa{} provides two routines to solve the ppRPA equation:
direct diagonalization and the Davidson algorithm\cite{yangExcitationEnergiesParticleparticle2014}.
Both routines use the density-fitting (resolution-of-identity) technique to compute electron repulsion integrals and can be combined with the active-space formalism\cite{liLinearScalingCalculations2023}.
For symmetry preserved spin-restricted ppRPA,
the ppRPA matrix is constructed in a spin-adapted manner\cite{yangBenchmarkTestsSpin2013}.
For spin-unrestricted ppRPA,
the ppRPA matrix is constructed and diagonalized in three subspaces
($\alpha, \alpha; \alpha, \alpha$),
($\alpha, \alpha; \beta, \beta$),
and ($\beta, \beta; \beta, \beta$)\cite{yangBenchmarkTestsSpin2013}.
As a lightweight package,
the core routines in \libpprpa{} only depend on standard Python packages such as Numpy\cite{harrisArrayProgrammingNumPy2020} and Scipy\cite{virtanenSciPy10Fundamental2020} for efficient mathematical operations,
which are parallelized with the OpenMP scheme.

After solving the ppRPA equation,
ppRPA two-electron addition/removal energies and eigenvectors in the numpy.ndarray format are returned as the output.
\libpprpa{} provides several analysis tools for ground-state and excited-state properties such as NTOs, density matrix and oscillator strengths.
For NTOs and density matrix,
the results can be saved in the Gaussian Cube format\cite{GaussianCubeFile} for further visualizations.
\libpprpa{} can also take the output to calculate analytic gradients for geometry optimizations,
which is integrated with \pyscf{} to obtained required integrals.

\FloatBarrier
\section{Example Usage}
\FloatBarrier
In this section the usage of \libpprpa{} for calculating excitation energies is demonstrated.
We begin with the particle-particle channel.
In Listing \ref{list:pp},
the calculation of charge-neutral excitation energies of \ce{H2O} in the particle-particle channel is shown.
The mean-field calculation of the ($N-2$)-electron system is performed by \pyscf{} first.
Then the input variables for ppRPA (the occupation number, orbital energies and the density-fitting matrix) are obtained from the mean-field object in \pyscf{}.
Finally the spin-adapted ppRPA calculations in the particle-particle channel are carried out.
\begin{lstlisting}[language=Python,caption={Example of using \libpprpa{} to calculate excitation energies of \ce{H2O} in the particle-particle channel.}, label={list:pp}]
from pyscf import dft, gto
# set up molecule
mol = gto.Mole()
mol.atom = [
    [8, (0.0, 0.0, 0.0)],
    [1, (0.0, -0.7571, 0.5861)],
    [1, (0.0, 0.7571, 0.5861)],
]
# ppRPA start with the (N-2)-electron system
# for hhRPA set mol.charge = -2
mol.charge = 2
mol.basis = "cc-pvdz"
mol.build()
# mean-field calculation
mf = dft.RKS(mol, xc="b3lyp")
mf.kernel()

from lib_pprpa.pyscf_util import get_pyscf_input_mol
nocc, mo_energy, R = get_pyscf_input_mol(mf)  # get input from PySCF

from lib_pprpa.pprpa_davidson import ppRPA_Davidson
# set channel="hh" for hhRPA
pprpa = ppRPA_Davidson(nocc, mo_energy, R, channel="pp")
pprpa.kernel("s")  # singlet calculation
pprpa.kernel("t")  # triplet calculation
pprpa.analyze()  #  analysis for excited states
\end{lstlisting}

Similar to the particle-particle channel,
the example for the calculation of charge-neutral excitation energies of \ce{H2O} in the particle-particle channel is shown in Listing \ref{list:pp}.
The mean-field calculation of the ($N+2$)-electron system is performed by \pyscf{} to generate input variables for \libpprpa{}.
Then the spin-adapted ppRPA calculations are performed in the hole-hole channel.

\FloatBarrier
\section{Results}
\FloatBarrier

\subsection{Computational Details}
\FloatBarrier
To demonstrate the performance of the \libpprpa{} library,
we reproduced a series of calculations in the literature:
S-T gaps of diradical systems\cite{yangSingletTripletEnergy2015},
double excitation of molecular systems\cite{yuAccurateEfficientPrediction2025},
charge transfer excitation of the Stein's set\cite{yangChargeTransferExcitations2017},
Rydberg excitation energies of atomic systems\cite{liLinearScalingCalculations2023},
valence excitation energies of the Thiel's set\cite{yangExcitationEnergiesParticleparticle2014},
vertical excitation energies of point defects\cite{liAccurateExcitationEnergies2024,liParticleParticleRandom2024},
reaction barriers in DBH24 set\cite{yangBenchmarkTestsSpin2013},
dissociation curves of \ce{H2} and \ce{Ar2}\cite{vanaggelenExchangecorrelationEnergyPairing2013},
potential energy surfaces of small molecules\cite{zhangAnalyticGradientsGeometry2014}.
The input SCF results and integrals for \libpprpa{} can be generated from any quantum chemistry package.
In this work,
we performed all ground-state SCF calculations in Gaussian basis sets with Gaussian density fitting using the \pyscf{} quantum chemistry software package~\cite{sunPySCFPythonbasedSimulations2018,sunRecentDevelopmentsPySCF2020}.

\FloatBarrier
\subsection{Excitation Energy}
\FloatBarrier

\begin{table}
\renewcommand{\arraystretch}{1.2}
\caption{Mean absolute errors (MAEs) of ppRPA based on HF, PBE and B3LYP for predicting different types of excitation energies.}\label{tab:excitation}
\begin{tabular}{c|ccc}
\toprule
                                & ppRPA@HF & ppRPA@PBE & ppRPA@B3LYP \\
\midrule
singlet-triplet gap (kcal/mol)  & 17.5     & 3.9       & 4.7         \\
double excitation (eV)          &          & 0.38      & 0.39        \\
charge-transfer excitation (eV) & 0.51     & 0.86      & 0.72        \\
valence excitation (eV)         & 0.79     & 0.39      & 0.37        \\
Rydberg excitation (eV)         & 0.08     & 2.13      & 2.53        \\
defect excitation (eV)          &          & 0.20      & 0.12        \\
\bottomrule
\end{tabular}
\end{table}

The mean absolute errors (MAEs) of ppRPA for predicting excitation energies of different characters,
including 24 singlet-triplet gaps of diradicals, 
29 double excitations in Loos's set\cite{kossoskiReferenceEnergiesDouble2024}, 
12 charge-transfer excitations in Stein's set\cite{steinPredictionChargetransferExcitations2009}, 
8 Rydberg excitations of atoms, 
38 valence excitations in Thiel's set\cite{silva-juniorBenchmarksElectronicallyExcited2008},
and 13 defect excitations are presented in Table~\ref{tab:excitation}.
It shows that ppRPA is capable with predicting accurate excitation energies for a broad range of systems and has a small starting point dependence on the chosen XC functional.
In particular,
for singlet-triplet gaps, double excitations and defect excitations,
ppRPA significantly outperforms TDDFT\cite{yangSingletTripletEnergy2015,liAccurateExcitationEnergies2024,liParticleParticleRandom2024,yuAccurateEfficientPrediction2025}.
In addition to the good accuracy,
the computational cost can be largely reduced by combining with the active-space formalism,
which is implemented in \libpprpa{} for both full diagonalization and Davidson algorithm.
As shown in Ref.\citenum{liLinearScalingCalculations2023},
ppRPA can be solved with a small active space without loss of accuracy.

\FloatBarrier
\subsection{Atomic Zero-Field Splitting}
\FloatBarrier

Relativistic ppRPA is a natural choice for calculating spin properties for systems with two open-shell systems.
We report the calculated atomic zero-field splittings (ZFS) for $^3P$ states of carbon group elements using X2CAMF-ppRPA method with uncontracted ANO-RCC basis sets  \cite{roosMainGroupAtoms2004}. Two-electron Coulomb interaction and Gaunt term have been included in X2CAMF Hamiltonian. All the calculations start with N-2 reference state. The occupied valence n$s$ orbital and virtual orbitals below 10 hartree are included in the ppRPA calculations.
The results are summarized in Table \ref{tab:zfs}. ppRPA@HF gives the best agreement with the experimental values. Relative errors are within 10\%. However, ppRPA@PBE and ppRPA@B3LYP tend to overestimate the ZFS by approximately 30\%. This might be due to the use of collinear functional.
\begin{table}
\renewcommand{\arraystretch}{1.2}
\caption{Calculated and experimental atomic zero-field splittings (ZFS) for $^3P$ states of carbon group elements using X2CAMF-ppRPA based on HF, PBE and B3LYP. The numbers are reported as the relative energy levels for J=1/J=2 states with respect to J=0 state.}\label{tab:zfs}
\begin{tabular}{c|cccc}
\toprule
ZFS (cm$^{-1}$) & ppRPA@HF & ppRPA@PBE & ppRPA@B3LYP & exp        \\
\midrule
C               & 14/42    & 21/62     & 20/59       & 16/43      \\
Si              & 73/211   & 104/300   & 99/287      & 77/223     \\
Ge              & 510/1295 & 806/1958  & 752/1842    & 557/1410   \\
Sn              & 1556/3212& 2429/4688 & 2279/4446   & 1692/3428  \\
Pb              &7191/10024&10671/14237& 10051/13518 & 7819/10650 \\
\bottomrule
\end{tabular}
\end{table}

\FloatBarrier
\subsection{Total Energy}
\FloatBarrier

\begin{figure}
\includegraphics[width=0.8\textwidth]{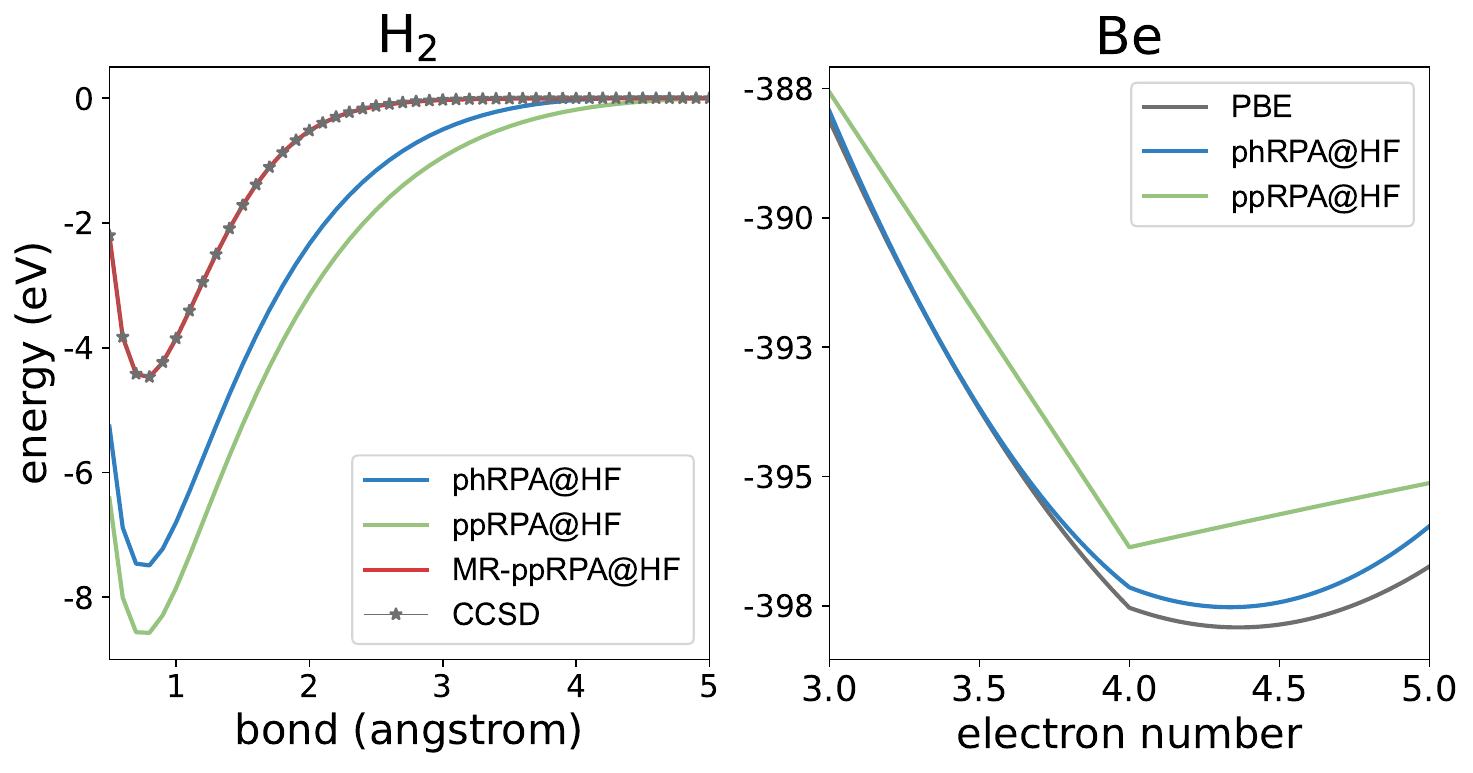}
\caption{Left: dissociation curve of \ce{H2} obtained from phRPA@HF, ppRPA@HF, ppRPA@HF (multireference DFT) and CCSD.
Right: behaviors of PBE, phRPA@HF and ppRPA@HF total energies of Be as a function of the electron number.
The cc-pVDZ basis set was used.}
\label{fig:total_energy}
\end{figure}

In this section,
the performance of \libpprpa{} for the total energy is shown.
Dissociation curves of single-bond breaking in \ce{H2} molecule and weakly interacting \ce{Ar2} molecules obtained from ppRPA are shown on the left-hand side of Fig.\ref{fig:total_energy}.
It shows that ppRPA@HF provides a similar dissociation curve to phRPA@HF for the \ce{H2} dissociation,
which overestimates the dissociation energies compared to the CCSD reference.
Because \ce{H2} only has two electrons,
MR-ppRPA is exact and gives the same results as the CCSD reference.
For other single-bond breaking problems,
MR-ppRPA treats the two valence electrons in a subspace configuration interaction fashion and is free from the static correlation error,
which has been shown to predict accurate dissociation energies and equilibrium bond lengths\cite{chenMultireferenceDensityFunctional2017,liMultireferenceDensityFunctional2022}.

On the right-hand side of Fig.\ref{fig:total_energy},
total energy obtained from ppRPA for fractional charge systems is shown.
In the context of DFT, 
ppRPA is the first known functional that captures the energy derivative discontinuity in strongly correlated systems.
As shown in Fig.\ref{fig:total_energy},
both phRPA and conventional functionals like PBE give convex curves for total energies between integer electron numbers,
which originate from large delocalization errors and lead to huge errors for predicting chemical potentials\cite{cohenInsightsCurrentLimitations2008,mori-sanchezFailureRandomphaseapproximationCorrelation2012}.
ppRPA has no delocalization error with a nearly linear energy behavior
for systems with fractional charges.
It captures the derivative discontinuity of the energy at integer electron numbers and meets the flat-plane condition\cite{mori-sanchezDiscontinuousNatureExchangeCorrelation2009},
which gives good accuracy for predicting ionization potentials and electron affinity using the finite-difference approach\cite{vanaggelenExchangecorrelationEnergyPairing2013}.

\FloatBarrier
\subsection{Analysis and Visualization}
\FloatBarrier

\begin{figure}
\includegraphics[width=0.4\textwidth]{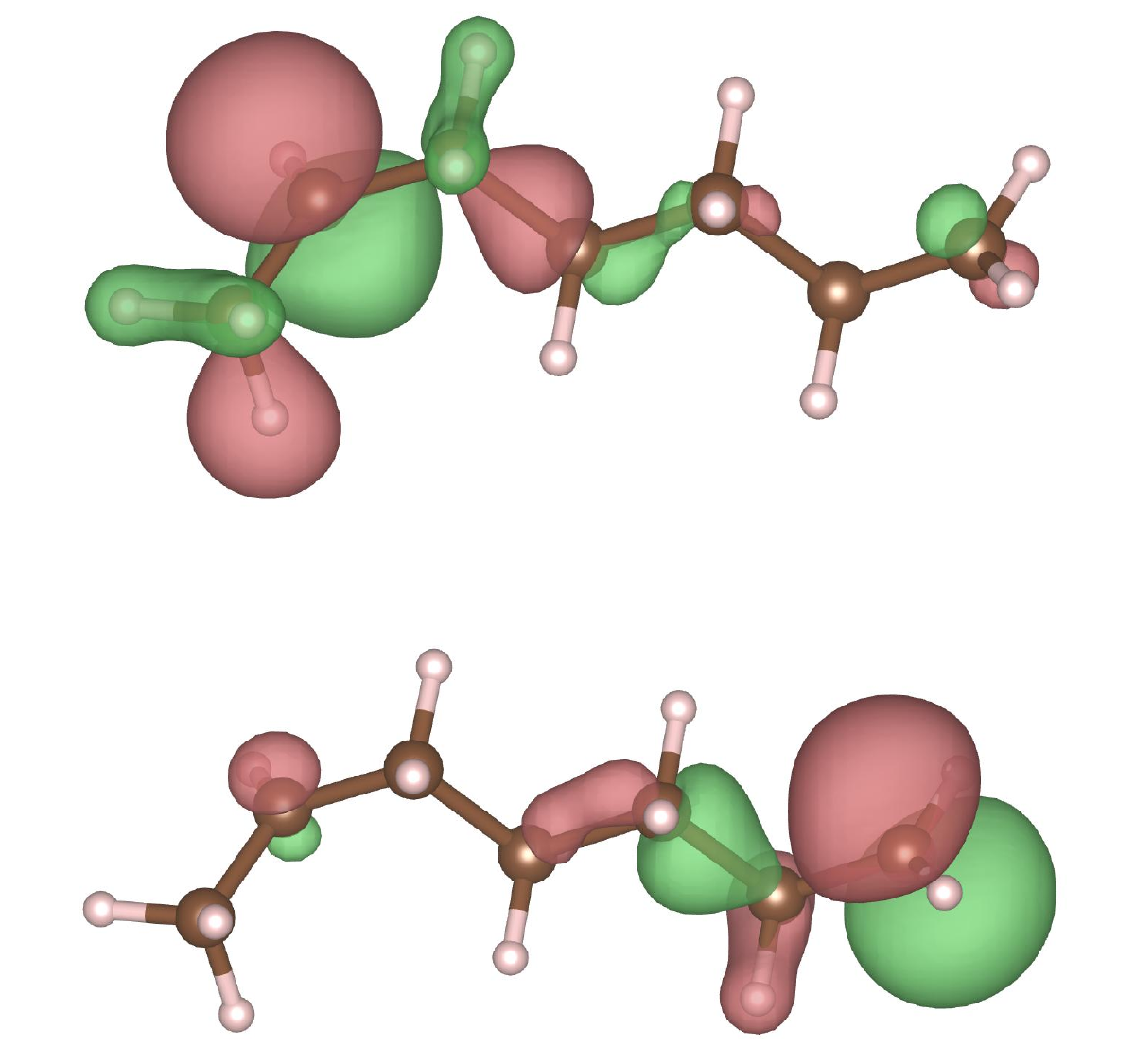}
\caption{Natural transition orbitals (NTOs) of the triplet ground state in $\cdot$\ce{CH2(CH2)4C(CH3)H}$\cdot$ obtained from ppRPA@B3LYP.
The NTO weight is 0.998.
Top: NTO of adding the first nonbonding electron.
Bottom: NTO of adding the second nonbonding electron.
The aug-cc-pVTZ basis set was used.
Geometry was taken from Ref.\citenum{yangSingletTripletEnergy2015}.
The isosurface value is 0.04 a.u.}
\label{fig:nto}
\end{figure}

\begin{figure}
\includegraphics[width=0.4\textwidth]{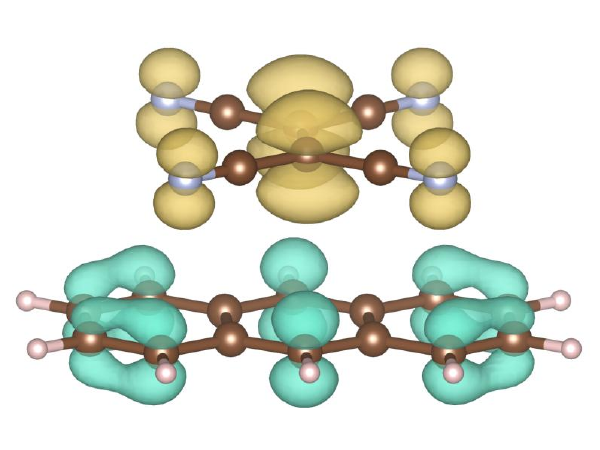}
\caption{Electron density difference between the first singlet excited state and the ground state of anthracene-TCNE obtained from ppRPA@B3LYP.
Yellow and blue indicate negative and positive electron densities, respectively.
The cc-pVDZ basis set was used.
Geometry was taken from Ref.\citenum{steinReliablePredictionCharge2009}.
The isosurface value is 0.003.}
\label{fig:dm}
\end{figure}

The ppRPA results can be further visualized and analyzed by NTOs and electron densities.
NTOs of the triplet ground state in $\cdot$\ce{CH2(CH2)4C(CH3)H}$\cdot$ obtained from ppRPA@B3LYP are shown in Fig.\ref{fig:nto}.
ppRPA provides good descriptions for this long disjoint diradicals.
It shows that two nonbonding electrons are added to carbon atoms on each chain end connected by a (\ce{CH2})$_n$ bridge.
The corresponding NTO weight is 0.998,
which means it can be properly described by a single-determinant approach and indicates the strong diradical character in $\cdot$\ce{CH2(CH2)4C(CH3)H}$\cdot$.

Then we visualize the charge transfer nature of the excitation in the anthracene-tetracyanoethylene (TCNE) system with the ppRPA density matrix.
The electron density difference between the first singlet excited state and the ground state of anthracene-TCNE obtained from ppRPA@B3LYP in Fig.\ref{fig:dm}.
It shows that the electron density flows from the aromatic donor anthracene to the TCNE acceptor.

\FloatBarrier
\subsection{Geometrical Gradient and Structural Optimization}
\FloatBarrier

\begin{figure}
\includegraphics[width=0.4\textwidth]{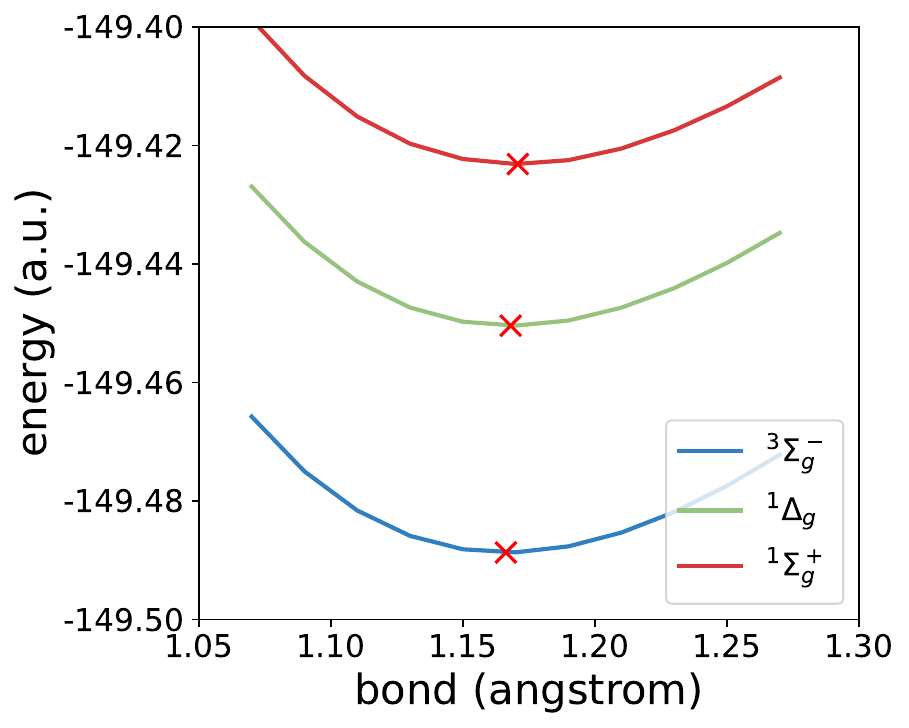}
\caption{Optimized structures for $^3\Sigma^-_g$, $^1\Delta_g$, and $^1\Sigma^+_g$ states of oxygen molecule together with the corresponding potential energy surfaces. The optimized structures are denoted as red cross. The equilibrium bond lengths are 1.1663, 1.1681, and 1.1708 \AA,  for  $^3\Sigma^-_g$, $^1\Delta_g$, and $^1\Sigma^+_g$ states, respectively.}
\label{fig:grad}
\end{figure}

With analytical gradient techniques \cite{zhangAnalyticGradientsGeometry2014}, \libpprpa{} can perform structural optimization using ppRPA methods. 
The $^3\Sigma^-_g$, $^1\Delta_g$, and $^1\Sigma^+_g$ states of oxygen molecule have been optimized to demonstrate the applicability of the ppRPA analytical gradient methods. ppRPA@HF method was used with cc-pVDZ basis sets and N-2 reference state. 
The optimized bond lengths and the corresponding potential energy surfaces are plotted in Fig. \ref{fig:grad}. 
The new implementation features several advances beyond the original publication \cite{zhangAnalyticGradientsGeometry2014}.
The details will be reported in the future work.

\FloatBarrier
\section{Conclusions}
In this work, 
we develop \libpprpa{}, 
an open-source and lightweight library 
for conducting ppRPA calculations. 
Implemented entirely in Python, 
\libpprpa{} provides a flexible and user-friendly framework 
that can seamlessly integrates with 
existing quantum chemistry packages 
such as \pyscf{}. 
By leveraging efficient matrix solvers, 
including direct diagonalization and the Davidson algorithm, 
the library enables accurate and scalable ppRPA calculations. 
Additionally, the incorporation of active-space approximations 
significantly reduces computational cost while maintaining high accuracy, 
making ppRPA more accessible for a wider range of molecular systems.

Through extensive benchmark studies, 
we have demonstrated the reliability and versatility of \libpprpa{} 
in predicting excitation energies, 
including S-T gaps, 
double excitations, 
charge-transfer excitations, 
and Rydberg excitations. 
Our results highlight the advantages of ppRPA 
over conventional particle-hole approaches, 
particularly in describing multireference character 
and strong correlation effects.

With its open-source nature and ease of integration, 
\libpprpa{} provides a robust platform for researchers 
to explore electronic excitations 
and advance the development of electronic structure methods. 
We anticipate that \libpprpa{} will serve as 
a valuable tool for the quantum chemistry community, 
facilitating both fundamental research 
and practical applications in excited-state calculations.

\FloatBarrier
\begin{acknowledgments}
J.Y. and W.Y. acknowledge the support from the National Institutes of Health (R35GM158181).
T.Z., J.L., and C. Z. are supported by the National Science Foundation (Grant No.~OAC-2513473). 
J.L. also acknowledges support from the Tony Massini Postdoctoral Fellowship in Data Science from Yale University. 
\end{acknowledgments}

\section*{Supporting Information}
See the Supporting Information for numerical results of excitation energies and ground-state energies obtained from ppRPA.

\section*{Data Availability Statement}
Data and scripts pertaining to this work have been archived in the Duke Research Data Repository\cite{dukedata}.

\bibliographystyle{apsrev4-2}
\bibliography{ref}

\end{document}